\documentclass[aps,prl,twocolumn,nofootinbib,amsmath,amssymb,a4paper]{revtex4}
\usepackage{graphicx}% Include figure files
\usepackage{dcolumn}% Align table columns on decimal point
\usepackage{bm}

\begin{document}

\title{Neutron $\beta$ Decay\\
Status and Future of the Asymmetry Measurement}% Force line breaks with \\

\author{Takeyasu M. Ito}
\affiliation{Los Alamos National Laboratory, PO Box 1663, MS H846, Los
 Alamos, NM 87545}
\email{ito@lanl.gov}

\begin{abstract}
With more intense sources of cold and ultracold neutrons becoming
available and with improved experimental techniques being developed,
determination of $|V_{ud}|$ from neutron $\beta$ decay with a similar
precision to that from from superallowed $\beta$ decays is within
reach. Determination of $|V_{ud}|$ from neutron $\beta$ decay, free
from nuclear corrections, hold the most promise for a further
improvement of the determination of $|V_{ud}|$. The current and future
neutron $\beta$ decay correlation experiments including the UCNA
experiment at Los Alamos National Laboratory are reviewed.
\end{abstract}

\maketitle

\section{Introduction}
High precision electroweak measurements provide stringent tests of the
standard model (SM) and search for what may lie beyond it. A
deviation from expectations based on our knowledge of the SM would be
indirect evidence for new physics. These high precision tests
complement direct searches for new physics using high energy collides.

Unitarity of the Cabibbo-Kobayashi-Maskawa matrix (CKM matrix)
requires that the first row satisfy
\begin{equation}
|V_{ud}|^2 + |V_{us}|^2 + |V_{ub}|^2 = 1.
\end{equation}
A deviation from unity could arise from additional heavy quark
mixing.  Also, because the extraction of these CKM matrix elements
involves comparison between muon decay and semileptonic decay of
hadrons, any new physics that causes a violation of quark-lepton
universality would cause a deviation from unity.

The value of $|V_{ud}|$ has been obtained from superallowed nuclear
$\beta$ decays, neutron $\beta$ decay, and pion $\beta$ decay. The
current determination of the value of $|V_{ud}|$ is dominated by
superallowed nuclear $\beta$ decays~\cite{HAR05}. The value is
(using an updated electroweak radiative correction~\cite{MAR06})
$|V_{ud}|=0.97377(11)(15)(19)$. The uncertainties are (1) nuclear
structure uncertainty added in quadrature with the experimental
uncertainty in the $ft$ value, (2) uncertainty in coulomb distortion
effects, and (3) uncertainty from quantum loop effects. Although the
third uncertainty is common to both $|V_{ud}|$ determination from
nuclear $\beta$ decay and that from neutron $\beta$ decay, $|V_{ud}|$
determination from neutron $\beta$ decay is free from nuclear
corrections that are associated with uncertainties (1) and
(2). Therefore, neutron $\beta$ decay can in principle provide a
determination of $|V_{ud}|$ with a smaller theoretical uncertainty
than nuclear $\beta$ decay.

With more intense sources of cold and ultracold neutrons becoming
available and with improved experimental techniques being developed,
determination of $|V_{ud}|$ from neutron $\beta$ decay with a similar
precision to that from nuclear $\beta$ decay is within reach. This
will provide a useful cross check of the current determination of
$|V_{ud}|$ from nuclear $\beta$ decays, in particular of the nuclear
dependent corrections. Furthermore, precision measurements of neutron
decay parameters hold the most promise for a further improvement of
the determination of $|V_{ud}|$.

Determination of $|V_{ud}|$ from neutron $\beta$ decay requires
knowledge of the neutron lifetime $\tau_n$ and the ratio of the axial
to vector coupling constants $\lambda = G_A/G_V$.  In the following,
after a brief review of the current status of the $|V_{ud}|$
determination, we will review the current and future neutron $\beta$
decay correlation measurements, which provide determination of
$\lambda$. The status of the neutron lifetime experiments is reviewed
in Ref.~\cite{BOW07}.

\section{Status of $V_{ud}$ Determination from Neutron $\beta$ Decay}
The value of $|V_{ud}|$ is determined by comparing the vector coupling
constant of $\beta$ decay, $G_V$, to the Fermi coupling constant,
$G_F$, determined from muon decay. In the case of free neutron $\beta$
decay, knowledge of both neutron lifetime $\tau_n$ and the ratio of
the axial to vector coupling constant $\lambda = G_A/G_V$ is required
to determine $G_V$.
\begin{eqnarray}
| V_{ud} |^2 & = & \frac{G_V^2}{G_F^2(1+\Delta_R)} \nonumber \\ 
             & = & \frac{2\pi^2}{G_F^2 m_e^5 \tau_n (1+3\lambda^2) f^R (1+\Delta_R)},
\end{eqnarray}
where $\Delta_R$ is the quantum loop correction mentioned earlier,
$m_e$ is the electron mass, $f^R$ is the phase space factor (including
the outer radiative correction). Numerically~\cite{MAR06},
\begin{equation}
| V_{ud} |^2 = \frac{4908.7(1.9)\;{\rm s}}{\tau_n (1+3\lambda)},
\end{equation}
where the uncertainty quoted is from $\Delta_R$.

The value of $\lambda$ is determined from measurements of decay
correlations.  The differential decay rate, averaged over electron
spin, is given by~\cite{JAC57}
\begin{eqnarray}
\lefteqn{\frac{dW}{dE_e d\Omega_e d\Omega_{\nu}}   \propto  p_e E_e 
\left ( E_0 - E_e \right )^2 } \nonumber \\
& &  \times  \left [ 1  + a\frac{\bm{p}_e\cdot\bm{p}_{\nu}}{E_eE_{\nu}} 
%  +b \frac{m_e}{E_e}  
 +  \left < \bm{\sigma}_n \right > \cdot 
\left (
A\frac{\bm{p}_e}{E_e}
+ B\frac{\bm{p}_{\nu}}{E_{\nu}}
%+ D\frac{\bm{p}_e\times \bm{p}_{\nu}}{E_eE_{\nu}}
\right )
\right ], 
\end{eqnarray}
where $m_e$ is the electron mass, $E_e$ the electron energy,
$\bm{p}_e$ the electron momentum, $E_{\nu}$ the neutrino energy,
$\bm{p}_{\nu}$ the neutrino momentum, and $\sigma_n$ the neutron
spin. Coefficients $a$, $A$, and $B$ depend only on $\lambda$ in the
SM.  Among them, $A$ is the most sensitive to $\lambda$ with
$\frac{d\lambda}{dA}=2.6$. $a$ has a similar but slightly reduced
sensitivity to $\lambda$ with $\frac{d\lambda}{da}={3.3}$. $B$ is much
less sensitive to $\lambda$ with $\frac{d\lambda}{dB}=13.4$. So far,
the determination of $\lambda$ from free neutron decay has been
provided by measurements of $A$. The uncertainty in $\lambda$ from the
most precise measurement of $a$~\cite{STR78} is more than ten times
larger than the uncertainty in $\lambda$ from the most precise
measurement of $A$~\cite{ABE02}. Therefore the main focus of this
review on is $A$ measurements.

\begin{figure}
\includegraphics[width=0.4\textwidth]{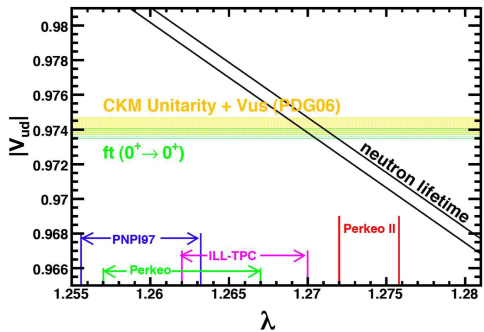}
\caption{ The current experimental status of the determination of
$|V_{ud}|$. The values of $\lambda$ from the four recent measurements
are shown by four brackets along the $\lambda$ axis as well as the
constraints from the world average of the neutron lifetime
measurements as given by the Particle Data Group~\cite{PDG2006}.
(Note that the average given by the Particle Data
Group~\protect\cite{PDG2006} does not include the recent measurement
reported in Ref.~\protect\cite{SER05}.)  The combination of a
$\lambda$ measurement and the neutron lifetime measurement determines
the value of $|V_{ud}|$. Also shown are the $|V_{ud}|$ determination
from nuclear $\beta$ decay and the $|V_{ud}|$ determination from kaon
and B-meson decays and the assumption of CKM unitarity.\label{fig:Vud}
}
\end{figure}

The current experimental situation is graphically summarized in
Fig.~\ref{fig:Vud}. The precision with which the recent four
measurements~\cite{ABE02,LIA97,YER97,BOP86} determined the value of
$A$ ($0.6-1.6$\%) is not sufficient to make a determination of
$|V_{ud}|$ from neutron $\beta$ decay with a precision comparable to
that from nuclear $\beta$ decay. Furthermore, the agreement among the
four measurements is poor. Clearly a new measurement of $A$ with a
higher precision is warranted.

A typical experimental arrangement for $A$ coefficient measurements
involves measuring the forward-backward asymmetry of electron emission
with respect to the neutron spin in polarized neutron $\beta$
decay. Polarized neutrons (a beam of cold neutrons in almost all
cases) are let decay in a decay volume and electrons from the neutron
decay are guided by a strong magnetic field towards one of the two
electron detectors located at the ends of the decay volume. When the
detectors have a $4\pi$ coverage of $\beta$ decay events, the
asymmetry in the count rate in the two detectors can be related to the
$A$ coefficient as follows:
\begin{equation}
\label{eq:A}
A_{\rm exp}(E_e) = \frac{N_1(E_e)-N_2(E_e)}{N_1(E_e)+N_2(E_e)} =
\frac{1}{2}PA\beta,
\end{equation}
where $E_e$ is the electron's energy, $N_{1(2)}$ is the count rate in
detector 1(2), $P$ is the average polarization of the neutrons, and
$\beta$ is the velocity of the electron in the units of the velocity
of light.

Three major sources of systematic uncertainties can be identified in
the previous experiments. They are (1) neutron polarization
determination, (2) background, and (3) detector effects including
backscattering of $\beta$ particles. As evident from Eq.~\ref{eq:A},
the polarization determination has to be done to a precision better
than the precision to which $A$ is to be determined. Also, incomplete
knowledge of the background signal will lead to an erroneous
determination of $N_{1(2)}$, thereby giving an erroneous determination
of $A$.  With regard to the detector effects, due to the small end
point energy of the electron spectrum ($E_e^0=782$~keV), a significant
fraction ($\sim 10\%$ for plastic scintillation counters) of electrons
from neutron $\beta$ decay directed to one detector can backscatter
from the surface of the detector and are detected by the other
detector. A non-negligible fraction of the backscattered electrons
leave undetectably small energy deposition in the first detector,
hence introducing an error in the asymmetry determination. (These
electrons are called missed backscattered electrons.) Understanding
the backscattering of low energy electrons and properly characterizing
the detector response is clearly of vital importance.

In order to address the unsatisfactory situation represented in
Fig.~\ref{fig:Vud}, measurements of $A$ with a precision of $\delta
A/A=0.2$\% or better are required (The uncertainty reported in
Ref.~\cite{ABE02} (PerkeoII experiment) is $\delta
A/A=0.6$\%). Clearly, these measurements need to address the
above-mentioned systematic issues. In Table~\ref{tab:systematics},
major systematic corrections applied to the results of the recent four
measurements are listed. It is seen that corrections that are
significantly larger than the reported uncertainty were
applied. Experiments with low background, high polarization
($>99.9$\%), and small detector effects are highly desirable since
they do not require large corrections, thus improving the reliability
of systematic error assignment.
\begin{table}
\caption{Major systematic correction in the recent $A$ measurements\label{tab:systematics} }
\begin{tabular}{cccc}\\ \hline\hline
Experiment & $A$ & \multicolumn{2}{c}{Systematic corrections} \\ 
{\ } & {\ } & Polarization & Background \\ \hline
Perkeo (1986) & -0.1146(19) & 2.6\% & 3\% \\ 
PNPI (1991) & -0.1116(14) & 27\% & small \\
ILL-TPC (1995) & -0.1160(15) & 1.9\% & 3\% \\
Perkeo II (2002) & -0.1189(7) & 1.1\%  & 0.5\%\footnote{After
  environmental   background(15\%) was subtracted.}\\ \hline\hline
\end{tabular}
\end{table}

\section{Ongoing and Future Neutron $\beta$-decay Correlation Experiments}
There are in fact several experiments ongoing or planned to measure
$A$ with a higher precision.

Since their last publication~\cite{ABE02}, Perkeo II collaboration
have implemented some upgrades, including a new ballistic supermirror
guide for a higher neutron flux~\cite{HAS02} and a new crossed
supermirror polarizers for a higher neutron
polarization~\cite{KRE05}. At the same time, a new experiment Perkeo
III has been developed.

There are two major efforts under way to measure $A$ in US. The UCNA
experiment~\cite{UCNA}, currently being commissioned at Los Alamos
National Laboratory, aims at a 0.2\% measurement of $A$ using
ultracold neutrons (UCNs). 

The abBA collaboration proposes to perform a simultaneous measurements
of $a$, $A$, $B$, and the Fierz interference term $b$ (which is zero
in the SM) at the Spallation Neutron Source (SNS)~\cite{WIL05}. The
goal of the abBA experiment is to determine $a$, $A$, $B$, and $b$
with an absolute precision of $\sim 10^{-4}$ In order to address known
problems in previous experiments, the abBA experiment includes several
new features such as, the use of pulsed neutron source, the use of a
polarized helium-3 transmission cell as a neutron polarizer,
coincidence detection of the decay electrons and the protons, and the
use of segmented silicon detectors. Since in the SM, $a$, $A$, and $B$
depends only on $\lambda$, the consistency among $a$, $A$, and $B$
will provide a powerful check for potential systematics.

There are also efforts to improve the precision of $a$. The aCORN
experiment, being prepared at NIST, aims to determine $a$ to a
statistical precision of 1\% or less by performing coincidence
detection of electrons and recoil protons and selecting two kinematic
regions such that a comparison of the rates in the two regions
directly yields a measurement of $a$~\cite{WIE05}. The aSPECT
experiment, currently being developed at Mainz and will be run at
ILL, will measure the recoil proton energy spectrum using a magnetic
spectrometer with electrostatic retardation
potentials~\cite{ZIM00}. The expected precision is $\delta
a/a=0.25$\%.

Below, we discuss the UCNA experiment more in detail.

\subsection{UCNA Experiment}
The goal of the UCNA experiment is a 0.2\% measurement of $A$. Unlike
previous experiments, which used a beam of cold neutrons from a
reactor, the UCNA experiment uses UCNs produced by a pulsed spallation
UCNs source~\cite{MOR02}. UCNs are neutrons with total kinetic energy
less than the effective potential $U_F$ presented by a material
boundary. These neutrons, therefore, can be confined in a material
bottle.  Typically $U_F\sim 200$~neV, which corresponds to velocities
of order 5~m/s, wavelengths of order 500~\AA\, and an effective
temperature of order 2~mK. 

There are two major advantages in using such a neutron source. First,
the kinetic energy of UCN is so small that the potential energy
associated with the interaction of the neutron magnetic moment with a
magnetic field ($\bm{\mu}\cdot \bm{B}$) can be easily made comparable
or even higher than the kinetic energy using an electromagnet that can
produce a magnetic field of several Tesla (1.7~T field gives
$|\bm{\mu}\cdot \bm{B}|=100$~neV ). Therefore, by passing UCNs through
a region with a large magnetic field ($>6$~T), it is possible to
filter out neutrons with one spin state, thereby making them 100\%
spin-polarized. Second, by operating the accelerator in a pulsed mode,
it is possible to limit the emission of background radiation to the
period in which the beam pulse strikes the target. By performing a
measurement only when there is no beam pulse striking the target, it
is possible to perform a measurement with very low background. This is
a big advantage of spallation sources over reactor sources, which
generate continuous background radiation.

\begin{figure}
\resizebox{7.5cm}{!}{\includegraphics{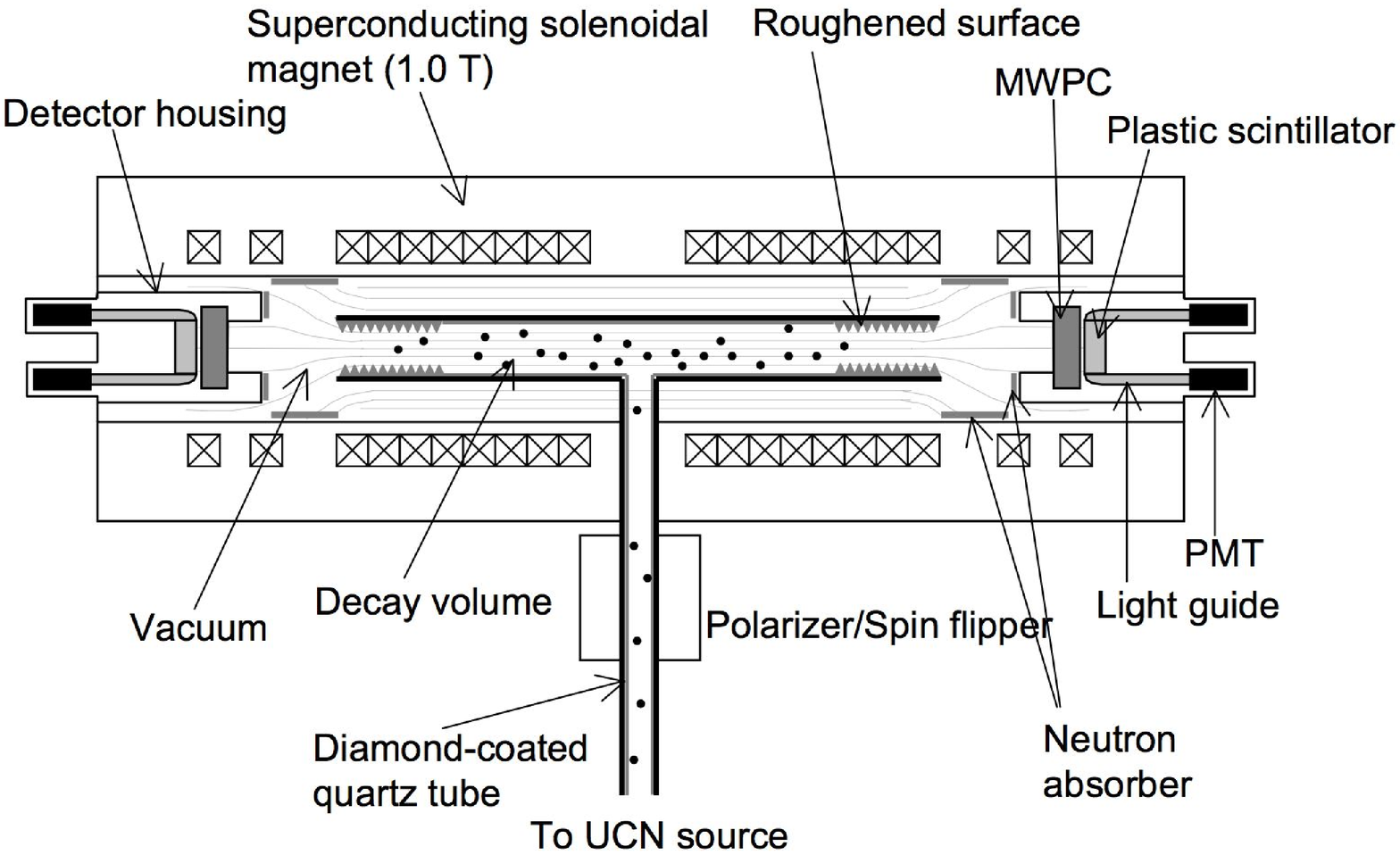}}
\caption{Schematic of the UCNA Experiment. }
\label{fig:ucna_schematic} 
\end{figure}

UCNs are produced by the LANSCE solid deuterium UCN source, sent
through a polarizer/spin flipper, and then introduced into a decay
volume. The wall of the decay volume is a 3~m-long diamond coated
quartz cylinder 10~cm in diameter.  The decay volume is in the warm
bore of a superconducting solenoidal magnet, which provides a holding
field of 1~T. The decay electrons spiral along the magnetic field
lines towards one of the detectors, and then enter the field expansion
region, where the magnetic field is reduced to 0.6~T. As an electron
enters the field expansion region, the energy associated with the
angular motion of the electron is transfered to longitudinal motion in
order to conserve angular momentum as the diameter of the spiral
increases due to the reduced field. This reduces the incident angle of
the electron onto the detector surface (reverse of the magnetic mirror
effect) and suppresses backscattering. The detectors are placed in a
region where the expanded field is uniform. The schematic of the UCNA
experiment is shown in Fig.~\ref{fig:ucna_schematic}.

Each electron detector is comprised of a thin low-pressure multiwire
proportional chamber (MWPC) placed in front of a plastic
scintillator. The MWPC consists of a thin front window (Kelvar
supported 6~$\mu$m-thick Mylar), a gas volume filled with neopentane,
cathode planes made of 50~$\mu$m diameter aluminum wires strung with a
2.54~mm spacing, an anode plane made of 10~$\mu$m diameter tungsten
wires strung with a 2.54~mm spacing, and a thin exit window
(6~$\mu$m-thick Mylar)~\cite{ITO07}. The MWPC provides the position
information with a resolution of $\sim 2$~mm, which is important in
rejecting $\beta$-decay events that occur near the wall of the decay
volume. Also the combination of the thin entrance window and the fact
that MWPCs are in general more sensitive to small energy depositions
than plastic scintillator reduces the fraction of missed backscattered
events. On top of this, detailed studies of low energy electron
backscattering were performed~\cite{MAR03} to help build a reliable
model of missed backscattered events, which will be necessary in
applying a small correction due to the missed backscattered events to
the final results. Furthermore, a small spectrometer was built to
provide a monoenergetic electron beam for off-line calibration of the
detector system~\cite{YUA01}.

\begin{figure}
\resizebox{7.5cm}{!}{\includegraphics{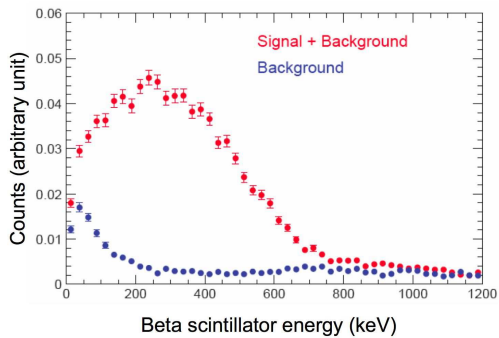}}
\caption{Spectrum of energy deposited in the plastic scintillation
counter\label{fig:beta} }
\end{figure}

Figure~\ref{fig:beta} shows an energy spectrum of the decay electrons
from UCNs that the collaboration obtained during their 2006
commissioning run. This is the first $\beta$ decay spectrum measured
with UCNs. The collaboration is hoping to make a $1-2$\% measurement
of $A$ during year 2007. 

\section{Summary and Outlook}
Determination of $|V_{ud}|$ from neutron $\beta$ decay with an
improved precision will provide a useful cross check of the current
determination of $|V_{ud}|$ from nuclear $\beta$ decays. Furthermore,
precision measurements of neutron decay parameters hold the most
promise for a further improvement of the determination of
$|V_{ud}|$. Currently there are a number of experiments ongoing and
planned that will determine $A$ and $a$ to a relative precision of the
order of 0.1\%, which, combined with a neutron life measurement with a
precision of 0.1\%, provide a determination of $|V_{ud}|$ with a
similar precision to that from nuclear $\beta$ decay.


\begin{thebibliography}{99}
\bibitem{HAR05}
J.~C.~Hardy and I.~S.~Towner, Phys. Rev. Lett. {\bf 94}, 092502
(2005); J.~C.~Hardy, these proceedings.

\bibitem{MAR06}
  W.~J.~Marciano and A.~Sirlin, Phys. Rev. Lett. {\bf 96}, 032002 (2006).

\bibitem{BOW07}
J.~D.~Bowman, these proceedings.

\bibitem{JAC57}
  J.~D.~Jackson, S.~B.~Treiman, and H.~W.~Wyld, {\sc Jr.},
  Phys. Rev. {\bf 106}, 517 (1957).

\bibitem{STR78}
C.~Stratowa, R.~Dobrozemsky, P.~Weinzierl, Phys. Rev. D {\bf 18}, 3970 (1978).

\bibitem{ABE02}
  H.~Abele {\it et al.}, Phys. Rev. Lett. {\bf 88}, 211801 (2002).

\bibitem{LIA97}
P. Liaud {\it et al.}, Nucl. Phys. A {\bf 612}, 53 (1997).

\bibitem{YER97}
B.~Yerozolimsky {\it et al.}, Phys. Lett. B {\bf 412}, 240 (1997).

\bibitem{BOP86}
P.~Bopp, {\it et al.}, Phys. Rev. Lett. {\bf 56}, 919 (1986).

\bibitem{PDG2006}
  W.-M.~Yao {\it et al.}, J. Phys. G {\bf 33}, 1(2006).

\bibitem{SER05}
A.~Serebrov {\it et al.}, Phys. Lett. B {\bf 605}, 72 (2005);
A.~P.~Serebrov {\it et al.} nucl-ex/0702009.

\bibitem{HAS02} 
H.~H\"{a}se, {\it et al.}, Nucl. Instrum. Methods Phys. Res. A
{\bf 485}, 453 (2002).

\bibitem{KRE05}
M.~Kreuz, {\it et al.}, Nucl. Instrum. Methods Phys. Res. A
{\bf 547}, 583 (2005). 

\bibitem{UCNA}
R.~Carr {\it et al.} (UCNA Collaboration), A proposal for an accurate
measurement of the neutron spin-electron angular correlation in
polarized neutron beta-decay with ultracold neutrons, 2000.

\bibitem{WIL05}
W.~S.~Wilburn, {\it et al.}, J. Res. Natl. Inst. Stand. Technol. {\bf
  110}, 389 (2005).

\bibitem{WIE05}
F.~E.~Wietfeldt, {\it et al.}, Nucl. Instrum. Methods Phys. Res. A
{\bf 545}, 181 (2005).

\bibitem{ZIM00}
O.~Zimmer, {\it et al.}, Nucl. Instrum. Methods Phys. Res. A {\bf
  440}, 548 (2000).

\bibitem{MOR02}
C.~L.~Morris {\it et al.}, Phys. Rev. Lett. {\bf 89}, 272501 (2002);
A.~Saunders {\it et al.}, Phys. Lett. B {\bf 593}, 55 (2004).

\bibitem{ITO07}
T.~M.~Ito {\it et al.}, Nucl. Instrum. Methods Phys. Res. A {\bf 571},
676 (2007).

\bibitem{MAR03}
J.~W.~Martin {\it et al.}, Phys. Rev. C {\bf 68}, 055503 (2003);
J.~W.~Martin {\it et al.}, Phys. Rev. C {\bf 73}, 015501 (2006).

\bibitem{YUA01}
J.~Yuan, {\it et al.}, Nucl. Instrum. Methods Phys. Res. A {\bf 465},
404 (2001).

\end{thebibliography}
\end{document}